\documentclass[aps,prl,twocolumn,groupedaddress]{revtex4}
\usepackage{graphicx}
\usepackage{graphpap}

\usepackage{dcolumn}
\usepackage{bm}

\begin{document}

\title{Neutralino Annihilation into Massive Quarks with SUSY-QCD Corrections}
\author{Bj\"orn Herrmann}
\affiliation{Institut f\"ur Theoretische Physik und Astrophysik,
 Universit\"at W\"urzburg,
 Am Hubland, D-97074 W\"urzburg, Germany}
\author{Michael Klasen}
\author{Karol Kova\v{r}\'{\i}k}
\email[]{kovarik@lpsc.in2p3.fr}
\affiliation{Laboratoire de Physique Subatomique et de Cosmologie,
 Universit\'e Joseph Fourier/CNRS-IN2P3/INPG,
 53 Avenue des Martyrs, F-38026 Grenoble, France}
\date{\today}
\begin{abstract}
We compute the full ${\cal O}(\alpha_s)$ supersymmetric (SUSY) QCD corrections for
neutralino annihilation into massive quarks through gauge or Higgs bosons and
squarks in the Minimal Supersymmetric Standard Model (MSSM), including the known
resummation of logarithmically enhanced terms. The numerical impact of the
corrections
on the extraction of SUSY mass parameters from cosmological data is analyzed for
gravity-mediated SUSY breaking scenarios and shown to be sizable, so that these
corrections must be included in common analysis tools.
\end{abstract}
\maketitle

\section{Introduction} 

\vspace*{-90mm}
\noindent LPSC 09-002\\
\vspace*{83mm}

The astrophysical and cosmological evidence gathered in recent years supports the
hypothesis of dark matter in our Universe. In particular, models of structure
formation favor Cold Dark Matter (CDM) consisting of Weakly
Interacting Massive Particles (WIMPs) with non-relativistic velocities.
The five-year data of the Wilkinson Microwave Anisotropy Probe (WMAP), combined
with the results of supernovae experiments and baryonic acoustic oscillation data,
constrain the relic density $\Omega$ of CDM in the Universe at 95\% ($2\sigma$) confidence
level to \cite{WMAP}
\begin{equation}\label{cWMAP}
  0.1097 < \Omega_{\rm CDM}h^2 < 0.1165\,.
\end{equation}
Here, $h$ denotes the present Hubble expansion rate $H_0$ in units of 100 km
s$^{-1}$ Mpc$^{-1}$.

In contrast to the Standard Model (SM) itself, its various extensions can provide
viable candidates for WIMPs. In the Minimal Supersymmetric Standard Model (MSSM)
with $R$-parity conservation, the Lightest Supersymmetric Particle (LSP) is such
a candidate, if it is a color singlet and electrically neutral. In
many scenarios, in particular those where SUSY breaking is mediated by gravity,
the LSP is the lightest neutralino and thus a suitable dark matter candidate in a
large part of the parameter space. To calculate the number density $n$ of the
relic particle, one has to solve the Boltzmann equation
\begin{equation}
  \frac{dn}{dt} ~=~ -3 H n - \langle \sigma_{\rm ann}v \rangle
  \big( n^2 - n_{\rm eq}^2 \big) ,
\end{equation}
where $n_{\rm eq}$ is the density of the relic particle in thermal equilibrium
and $v$ is the relative velocity of the annihilating pair. 

The thermally averaged annihilation cross section $\langle \sigma_{\rm ann}v
\rangle$ includes all (co-)annihilation processes of the dark matter particle into
SM particles. It is dominated by two-particle final states, most notably by
fermion-antifermion pairs and by combinations of gauge ($W^\pm,Z^0$) and Higgs
bosons ($h^0,H^0,A^0,H^\pm$) \cite{JungmanDrees}. The nature of the final state
contributing most depends strongly on the region of parameter space. Fermion final
states have the clear advantage that they are always kinematically allowed. Their
leading contribution is proportional to the mass of the fermion, so that we focus
our attention here on the annihilation into the massive quarks of the third
generation.

The relic density of the lightest neutralino depends on the SUSY-breaking
parameters of the MSSM, which determine the nature of the neutralino as well as
the couplings and masses that appear in the annihilation cross section. The number
of free parameters is often reduced to a few universal parameters
imposed at the high-scale e.g. the five parameters $m_0$, $m_{1/2}$, $A_0$,
$\tan\beta$, and sgn($\mu$) in minimal supergravity (mSUGRA). Using the
experimental limits in Eq.~(\ref{cWMAP}), one can then constrain the parameter
space in a complementary way to collider and low-energy experiments.

This analysis is made possible by public computer codes, which perform the
calculation of the dark matter relic density within models beyond the SM. The most
popular and developed codes are {\tt DarkSUSY} \cite{Dsusy} and {\tt micrOMEGAs}
\cite{microm}. The implemented processes are mostly calculated at leading order,
and higher order corrections are only included for some very sensitive
quantities. However, owing to the large magnitude of the strong coupling constant,
all QCD and SUSY-QCD corrections significantly affect the annihilation
cross section.

In this Letter, we extend our previous work \cite{lett1} by computing the full QCD
and SUSY-QCD corrections to neutralino annihilation into top and bottom
quark-antiquark pairs. As an example, we present the impact of the radiative
corrections in mSUGRA scenarios where the annihilation into top quark pairs is
dominant. Moreover, we include a study of bottom quark final states at small
$\tan\beta$, where we find agreement with previous results of a similar
calculation \cite{Baro:2007em}. Since light quark final states do not lead to sizable contributions in the analyzed mSUGRA
scenarios, their discussion is postponed to a later publication.

\section{Calculation details}

\begin{figure}[t!]
	\includegraphics[scale=0.57]{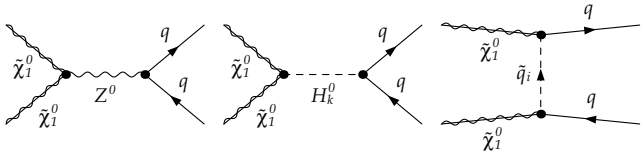}
	\caption{Tree-level diagrams for the annihilation of neutralinos into quarks. The crossed $u$-channel diagram is not shown. We denote the neutral Higgs bosons $(h^0,H^0,A^0)$ by $H_k^0$.}\label{FDtree}
\end{figure}

The annihilation of neutralinos into quarks proceeds at tree-level through an
exchange of the $Z$-boson and Higgs-bosons in the $s$-channel or of scalar quarks
(squarks) in the $t$- and $u$-channels (see Fig.~\ref{FDtree}). To reach a sufficient annihilation rate,
the cross section has to be enhanced either by an $s$-channel resonance or by a
small squark mass.

\begin{figure}[t!]
\begin{picture}(250,200)
 \put(0,145){\includegraphics[scale=0.57]{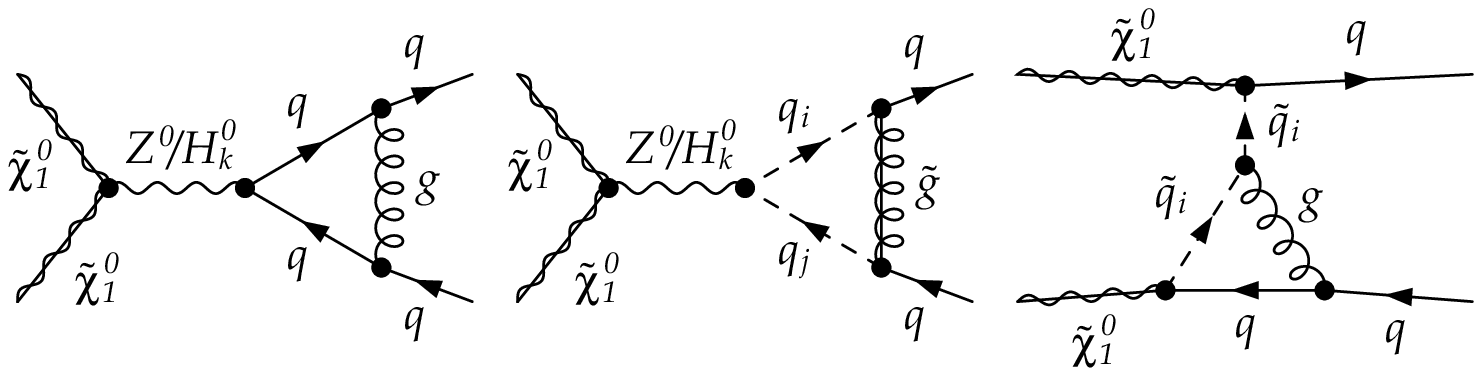}}
 \put(0,70){\includegraphics[scale=0.57]{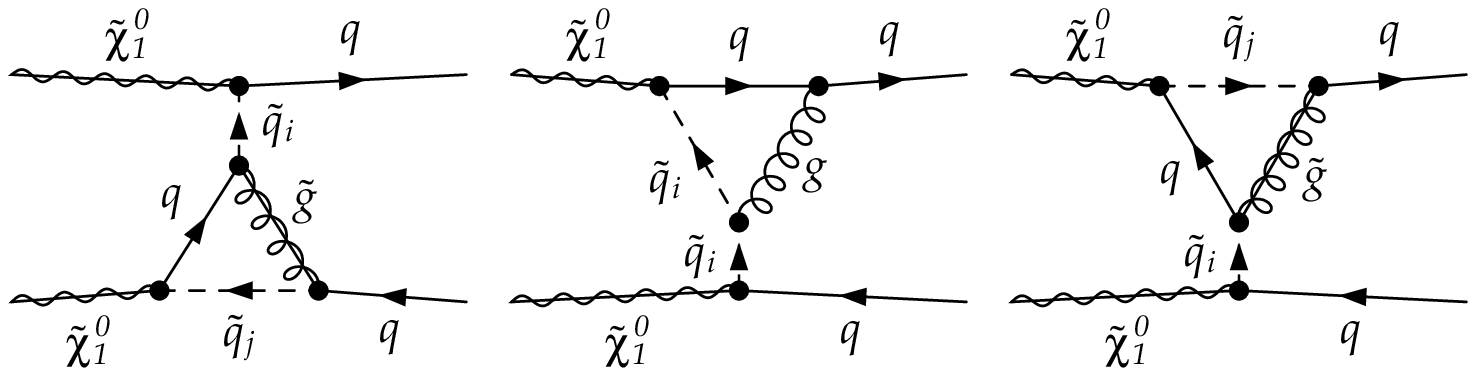}}
 \put(0,5){\includegraphics[scale=0.57]{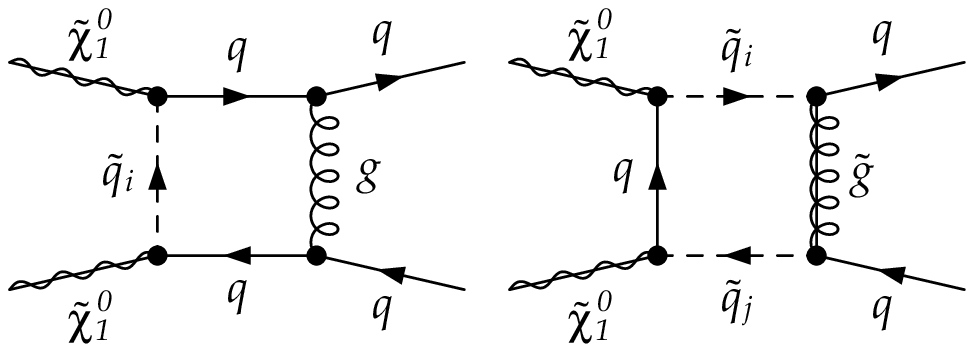}}
\end{picture}
	\caption{SUSY-QCD loop diagrams contributing to the annihilation of neutralinos into quarks. We leave out the diagrams for the $u$-channel contributions and the crossed box diagrams.}\label{FDloop}
\end{figure}

For these processes, we calculate the full one-loop QCD and SUSY-QCD corrections
involving the exchange of a gluon or a gluino. These affect the vertices including
$Z$- and Higgs-bosons and a quark-antiquark pair as well as those with
neutralinos, quarks, and squarks. In addition, one has to include the box
diagrams arising from the exchange of a gluon or a gluino between the final state
quarks in the $t$- and $u$-channels (see Fig.~\ref{FDloop}). All relevant divergent integrals are
evaluated in the modified dimensional reduction scheme ($\overline{\rm DR}$). We
use on-shell renormalization of the wave functions and couplings in order to
eliminate the ultraviolet divergencies. The squark mixing matrix is also
renormalized on-shell as proposed in Ref.\ \cite{EberlKovarik}. In the Yukawa
couplings, however, we use the quark masses defined in the $\overline{\rm DR}$
renormalization scheme. Full details of the calculation will be presented in a
forthcoming paper.

For the bottom quark, we start from the input value $m_b^{\overline{\rm MS}}
(m_b)$, evolve to the scale $Q$ using three-loop SM renormalization group
equations, change from $\overline{\rm MS}$ to $\overline{\rm DR}$ scheme
using the
corresponding relation at two loops \cite{Baer:2002ek}, and finally include the
MSSM threshold corrections comprising the sbottom-gluino and stop-chargino
one-loop contributions. The latter are considerably enhanced for large $\tan\beta$
or large $A_b$, so that they have to be resummed to all orders of perturbation
theory \cite{CarenaSpira}. Denoting the resummable part by $\Delta_b$ and the
finite one-loop remainder by $\Delta m_b$, the bottom quark mass is then given by
\begin{equation}
  m_b^{\rm MSSM}(Q) ~=~ \frac{m_b^{\overline{\rm DR}}(Q)}{1+\Delta_b} -
  \Delta m_b\, .
\end{equation}

The $\overline{\rm DR}$-mass of the top quark is obtained from the on-shell value
$m_t=172.4\ {\rm GeV}$ \cite{CDFD02008vn} by subtracting the finite part of the
on-shell mass counterterm. 

The infrared divergence connected to the exchange of a massless gluon is also
regularized dimensionally. To cancel the infrared divergent poles, we include the
bremsstrahlung process with an additional gluon in the final state. To allow for
integration of the divergent bremsstrahlung matrix element and for the
cancellation of the infrared divergencies, we use the dipole subtraction formalism
for massive partons \cite{Catani:2002hc}.   

We have performed several checks of the calculation. In particular, we have
compared our analytical calculation of the virtual part with the one produced by
the packages {\tt FeynArts} and {\tt FormCalc} \cite{FA} and the QCD corrections
to the $s$-channel Higgs-boson exchange with the known results of Ref.\
\cite{DreesBraaten}.

\begin{table}[t!]
  \begin{center}
    \begin{tabular}{c|ccccc|c|cc|}
       & $m_0$ & $m_{1/2}$ & $A_0$ & $\tan\beta$ & sgn$(\mu)$ & $\Omega_{\rm CDM}h^2$ & $b\bar{b}$ & $t\bar{t}$ \\
       \hline
       1 & 1800 & 131 & -1500 & 10 & + & 0.116 & 86\% & -- \\
       2 & 5800 & 536 & -1500 & 10 & + & 0.111 & -- & 72\% \\
       3 & 3200 & 520 & 0 & 50 & + & 0.110 & 8\% & 71\% \\
       \hline
    \end{tabular}
  \end{center}
  \vspace*{-4mm}
  \caption{High scale mSUGRA parameters (in GeV) together with the corresponding
  neutralino relic density and the contributions of the massive quark anti-quark
  final states to the annihilation cross section for our selected scenarios.}
  \label{Tpoints}
\end{table}

\section{Numerical analysis}

To evaluate the impact of the QCD and SUSY-QCD corrections on the neutralino relic
density, we have chosen three typical parameter points in mSUGRA shown in
Tab.~\ref{Tpoints}. They were selected so that the top and bottom quark anti-quark
final states dominate the total annihilation cross section and the relic density
lies within the WMAP range of Eq.~(\ref{cWMAP}). Moreover, the SUSY particle mass
spectrum resulting from the high scale parameters respects the experimental mass
exclusion limits, and the branching ratio of the decay $b\to s\gamma$ lies within
$2\sigma$ of the experimental bound 
\begin{math}
	{\rm BR}(b \to s \gamma) ~=~ \big( 3.52 \pm 0.25 \big) \cdot 10^{-4}\,,	
\end{math}
obtained from combined measurements by BaBar, Belle, and CLEO \cite{bsgamma}.
  
\begin{figure}[t!]
  \begin{center}
    \includegraphics[scale=0.32]{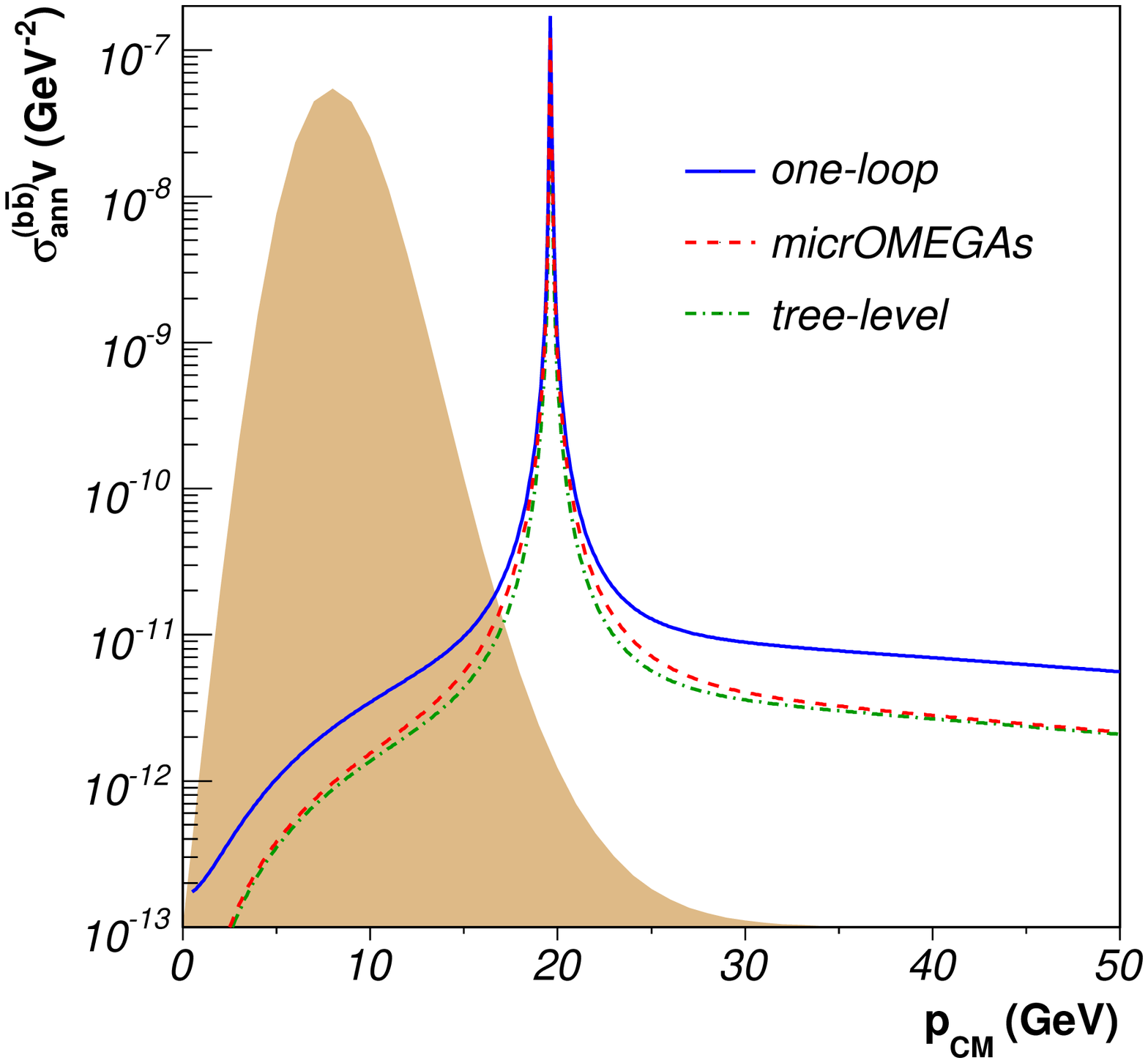}\vspace*{-3.5mm}
    \includegraphics[scale=0.32]{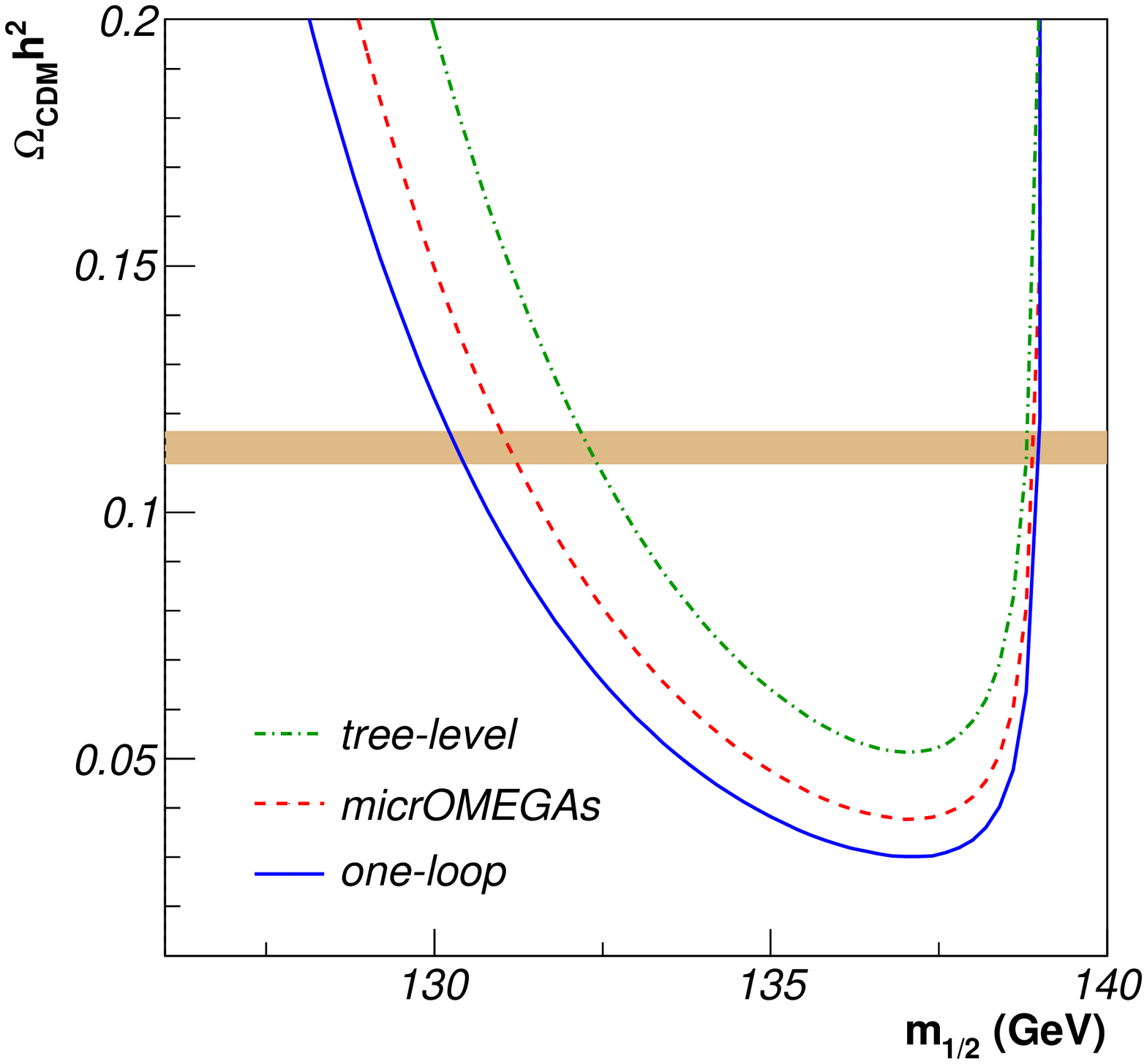}\vspace*{-3.5mm}
    \includegraphics[scale=0.32]{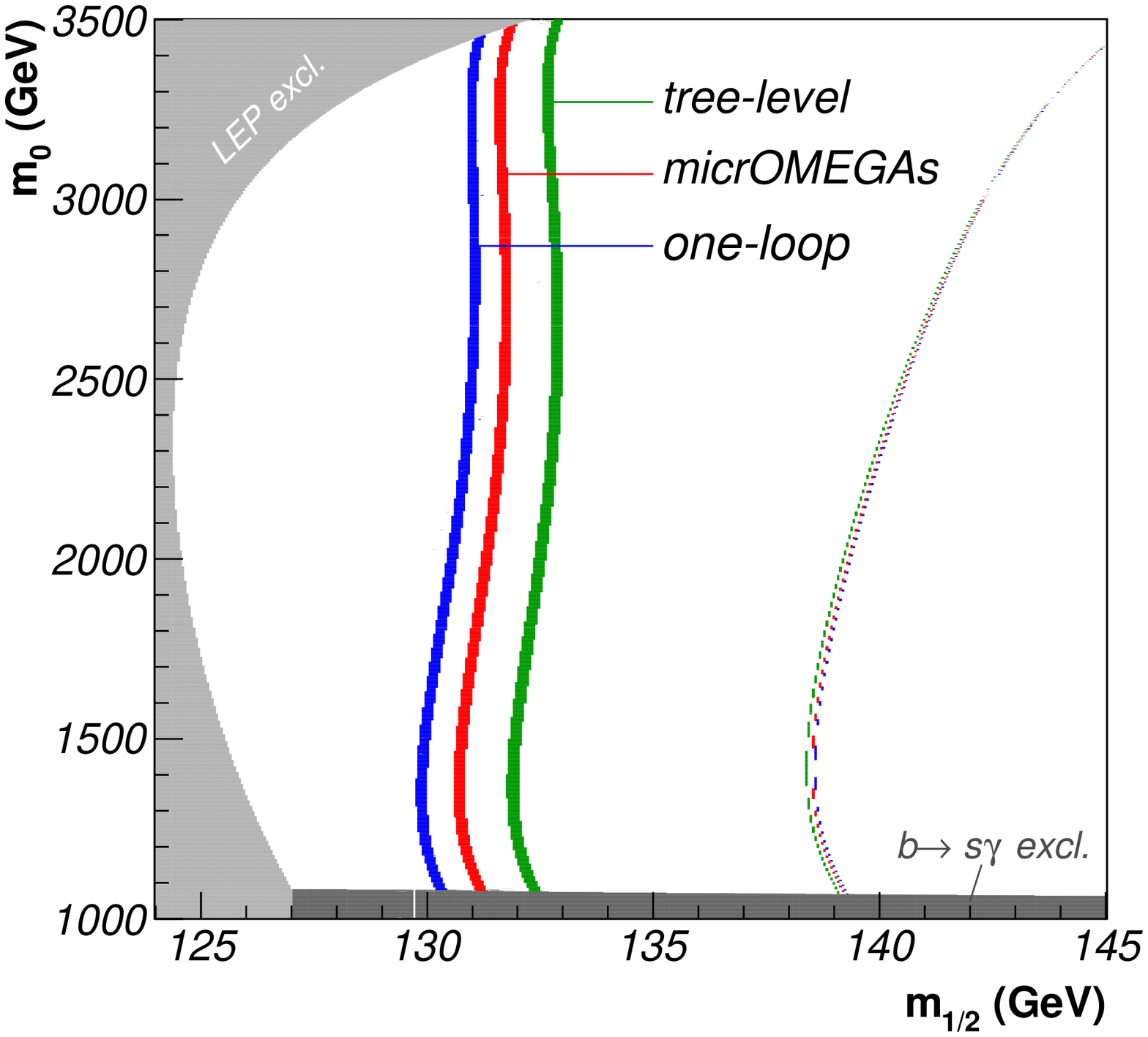}
  \end{center}
  \vspace*{-8mm}
  \caption{The effects of the radiative corrections on the cross section as a
  function of center-of-mass momentum (top), the prediction of the neutralino
  relic density as a function of the gaugino mass parameter $m_{1/2}$ (center),
  and on the cosmologically favored regions in the $m_0$--$m_{1/2}$ plane
  (bottom) for our parameter point 1 with annihilation into bottom quarks through the light  
  Higgs boson resonance.}
\label{Fig1}
\end{figure}

The high-scale parameters are evolved down to the electroweak scale using
{\tt SPheno 2.2.3} \cite{spheno}. The neutralino relic density is calculated with
{\tt micrOMEGAs 2.1}, where we have included the full radiative QCD and SUSY-QCD
corrections to the annihilation cross section. Our first parameter point at low
$\tan\beta$ has been chosen near the bulk region with a low fermion mass parameter
$m_{1/2}$, but a rather large scalar mass $m_0$ in order to avoid the constraint
coming from BR($b\to s \gamma$). The remaining top-quark dominated points lie both
in the focus point region, where $m_0$ is very large. At large $\tan\beta$, there
is also a contribution from bottom quarks due to their important coupling to the
$CP$-odd Higgs boson. 

\begin{figure}[t!]
  \begin{center}
    \includegraphics[scale=0.32]{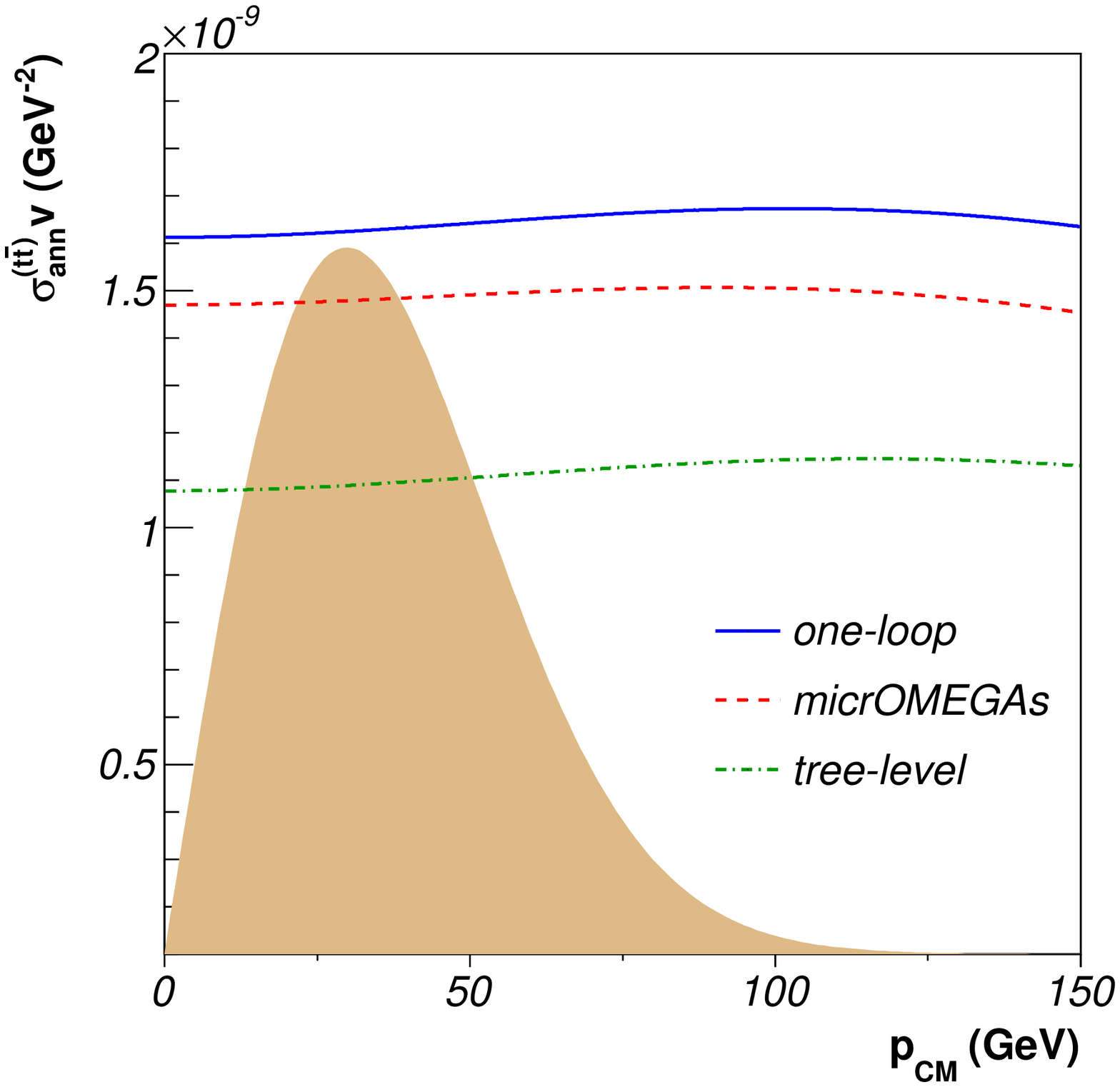}\vspace*{-3.5mm}
    \includegraphics[scale=0.32]{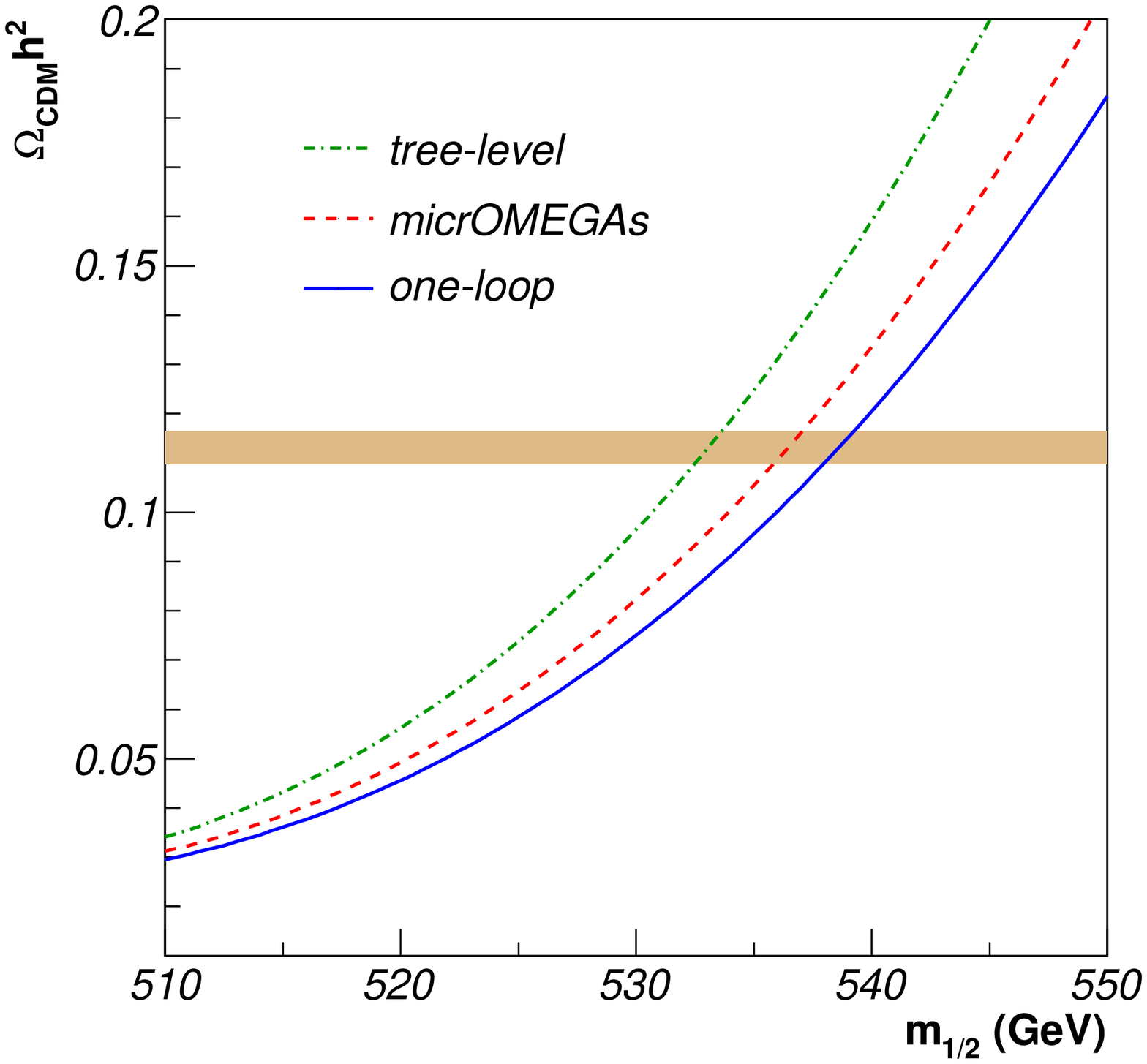}\vspace*{-3.5mm}
    \includegraphics[scale=0.32]{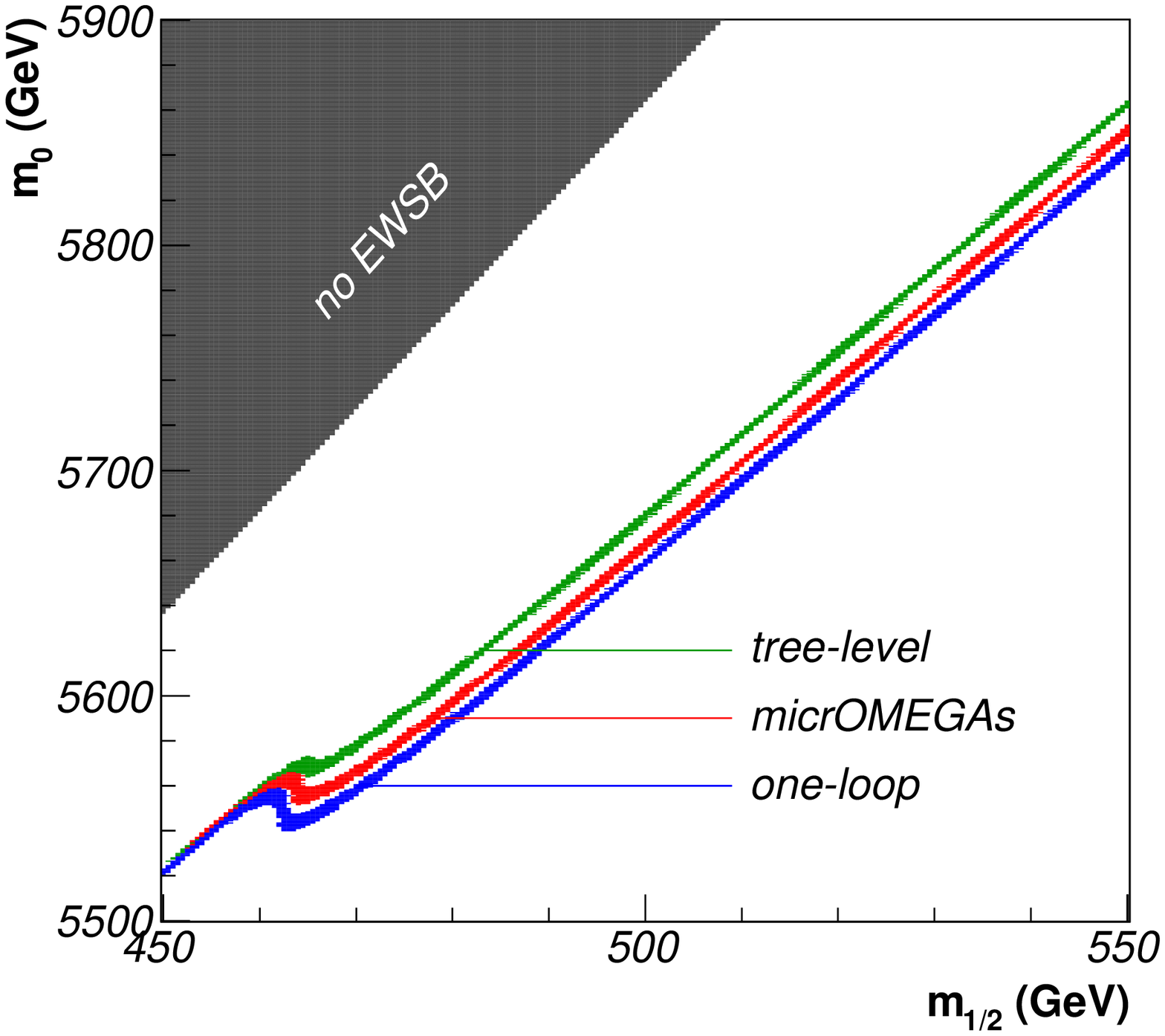}
  \end{center}
  \vspace*{-8mm}  
  \caption{Same as Fig.\ \ref{Fig1} for our scenario 2 with neutralino annihilation dominantly into top quarks.}	
\label{Fig2}
\end{figure}

\begin{figure}[t!]
  \begin{center}
    \includegraphics[scale=0.32]{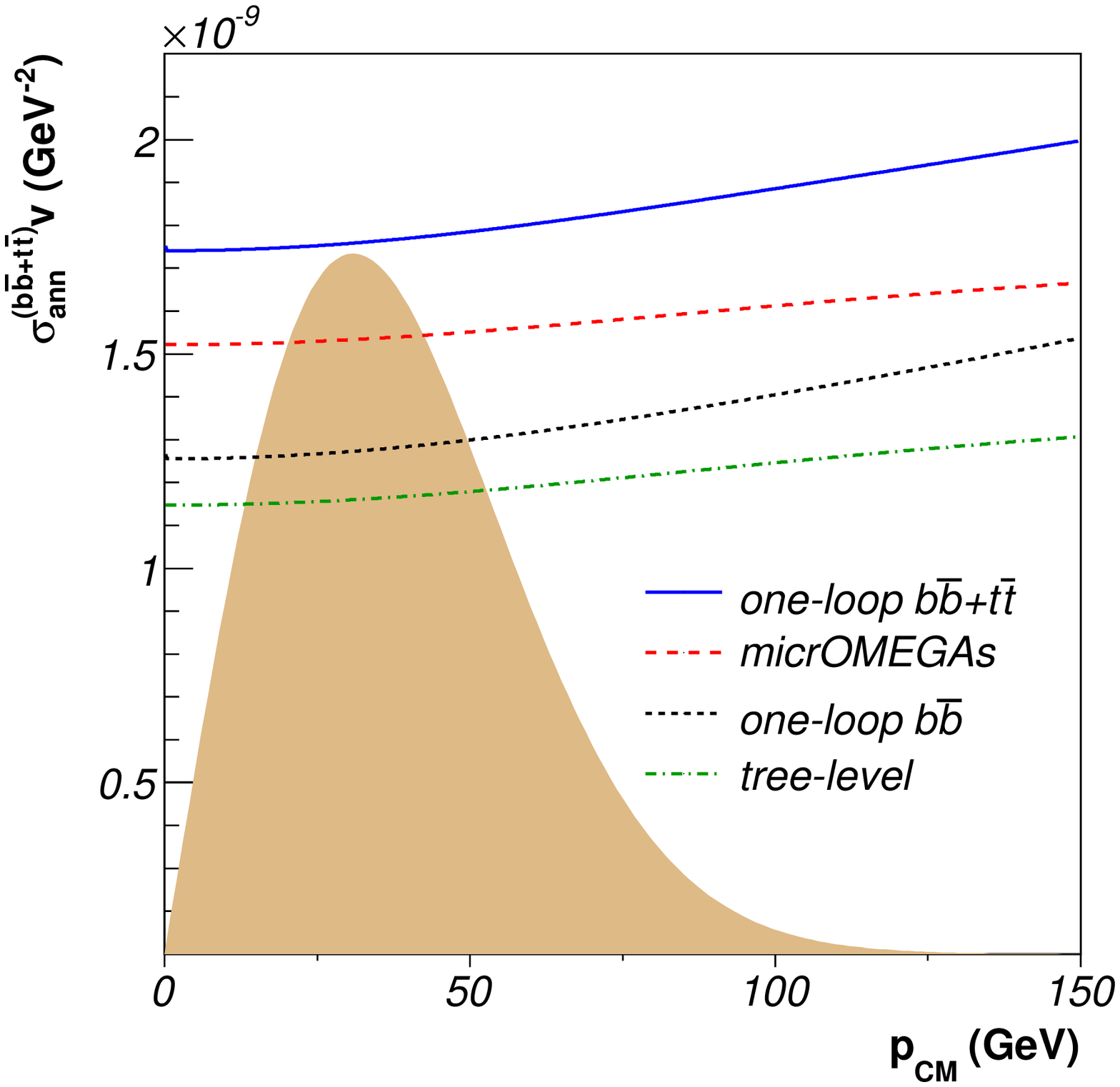}\vspace*{-3.5mm}
    \includegraphics[scale=0.32]{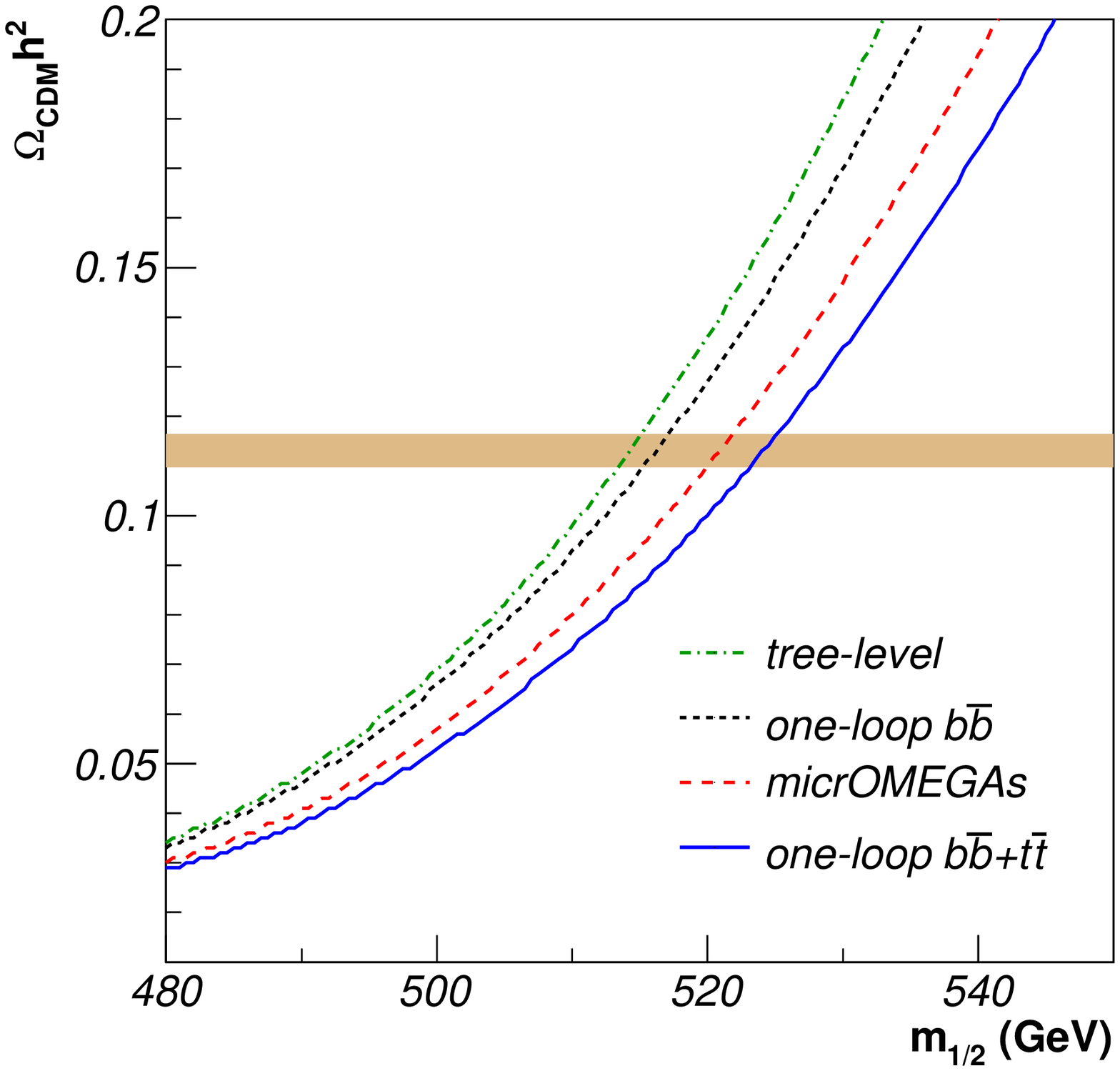}\vspace*{-3.5mm}
    \includegraphics[scale=0.32]{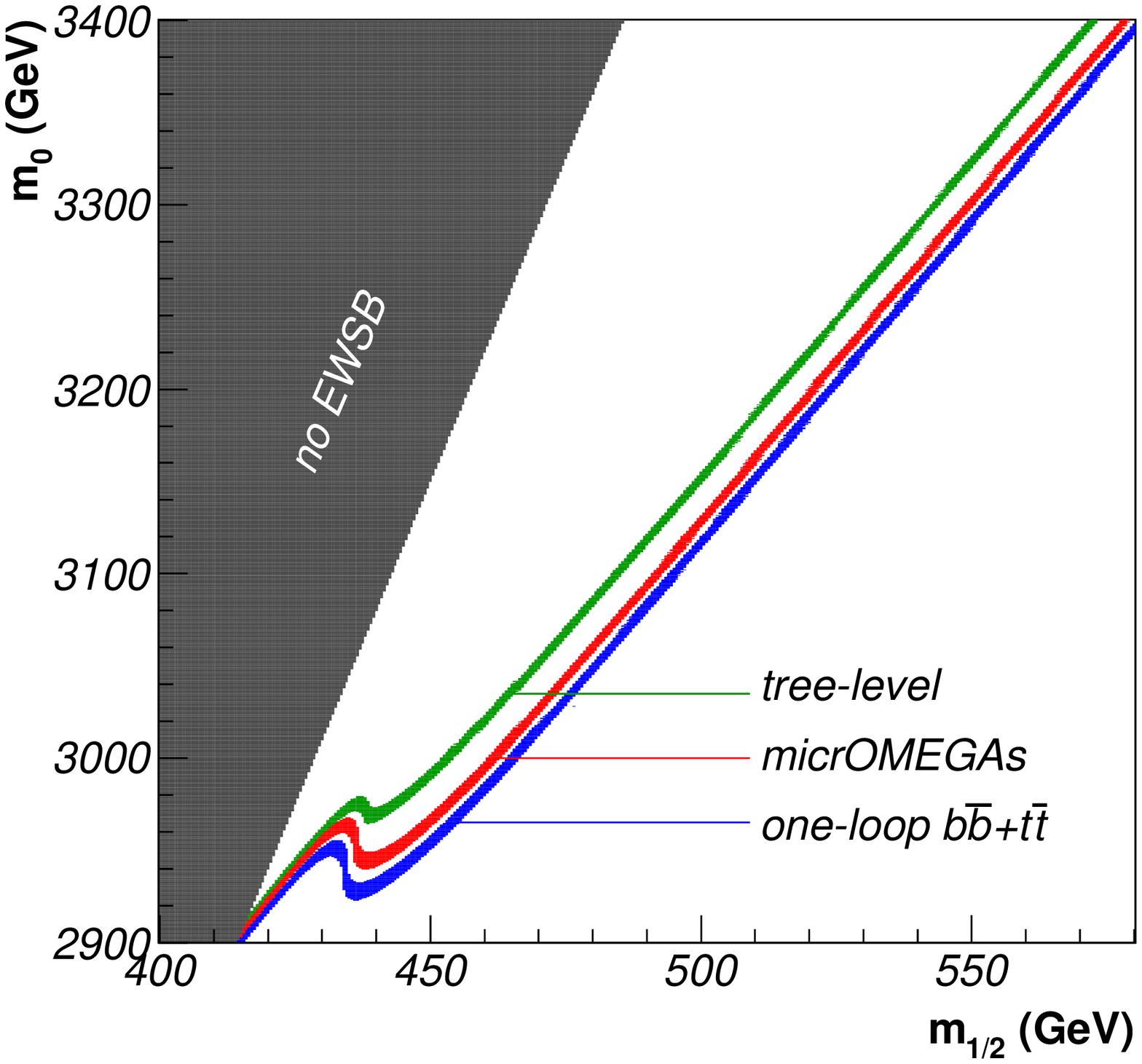}
  \end{center}
  \vspace*{-8mm}
  \caption{Same as Fig.\ \ref{Fig2} for our scenario 3 with neutralino
  annihilation into top and bottom quarks. The corrections to the $b\bar{b}$ final
  state only are also shown separately.}	
\label{Fig3}
\end{figure}

In the first panels of Figs.~\ref{Fig1}--\ref{Fig3}, we show the numerical
results for the relevant annihilation cross sections as a function of the relative
momentum $p_{\rm cm}$, that is related to the center-of-mass energy $\sqrt{s}$
and the neutralino mass $m_{\tilde{\chi}}$ through $s=4 \big(p_{\rm cm}^2+
m_{\tilde{\chi}}^2 \big)$. Shown are the leading order results with
$\overline{\rm DR}$ Yukawa couplings (dash-dotted), the approximation already
included in {\tt micrOMEGAs} (dashed), and the result including our full one-loop
QCD and SUSY-QCD corrections (solid). We also show, in arbitrary units, the
Boltzmann distribution function involved in the calculation of the thermal average
at the freeze-out temperature (shaded area). It indicates which centre-of-mass momenta contribute most to the relic density. E.g. in Fig.~\ref{Fig1} one sees that an important contribution comes also from energies below the light Higgs boson resonance, which is clearly visible on the top panel. 


The center panels of Figs.\ \ref{Fig1}--\ref{Fig3} show the corresponding
predictions for the neutralino relic density as a function of the gaugino mass
parameter $m_{1/2}$ around our parameter points 1, 2, and 3, respectively. The
horizontal bands indicate the experimental limits of Eq.\ (\ref{cWMAP}). Due to
the enhanced cross section, the relic density is reduced by about the same amount,
and the region of parameter space agreeing with the constraints is modified. 

In the bottom panels we finally show the cosmologically favored regions in the
$m_0$--$m_{1/2}$ plane that result from a relic density calculation using the
cross section at the tree-level, the approximation included in {\tt micrOMEGAs},
and the full one-loop SUSY-QCD corrected cross section. Note that the effect of
the corrections is larger than the experimental errors, so that three distinct
bands are observed. The grey shaded regions correspond to the points that are
excluded due to mass limits, no electroweak symmetry breaking (EWSB), or the
constraint from ${\rm BR}(b\to s\gamma)$. Concerning the ``bulk'' region at low
$\tan\beta$, the favored band is shifted to smaller values of $m_{1/2}$ and thus
towards the region excluded by LEP mass limits. The cosmologically favored focus
point region, however, is found at higher values of $m_{1/2}$ and lower values of
$m_0$ after including the SUSY-QCD corrections.

Our additional corrections to the relic density for the case shown in Fig.~\ref{Fig1} amount to about 10\% with respect to the approximation implemented in {\tt micrOMEGAs}. This is due to the low value of $\tan\beta$, which suppresses the part of the usually dominant resummed correction which is included in {\tt micrOMEGAs}, making the rest of the SUSY-QCD correction relevant. In the case of top quark final states, shown in Figs.~\ref{Fig2} and \ref{Fig3}, the additional corrections can amount up to 15\%. Note that SUSY-QCD corrections for top quark final states are not considered in {\tt micrOMEGAs}. It is interesting that in the focus point scenario with large $\tan\beta$, the correction to the bottom quark final state has a sizable contribution to the total correction. As a result, the relative contribution of the annihilation into bottom quarks increases compared to the 8\% at tree-level. This is well visible in the centre and bottom panels of Fig.~\ref{Fig3} and shows the importance of the $\tan\beta$ enhanced corrections to the bottom Yukawa coupling.

\section{Conclusions}

In summary, we have calculated the full one-loop QCD and SUSY-QCD corrections to
neutralino annihilation into third generation quarks. A numerical evaluation has
shown that the corrections have sizeable effects on the annihilation cross section
and in consequence on the extraction of SUSY mass parameters from cosmological
data assuming that the lightest neutralino is the cold dark matter candidate. The
induced difference is of the same order of magnitude as the experimental error
from cosmological precision measurements, so that the full one-loop corrections
should be taken into account when analyzing the SUSY parameter space with respect
to cosmological data.

\begin{acknowledgments}
The authors would like to thank A.~Pukhov for his help in implementing the results
into the {\tt micrOMEGAs} code. The work of K.K.\ is supported by the ANR project
ANR-06-JCJC-0038-01 and the work of B.H.\ is supported by DFG-Project PO 1337/1-1
and by a MENRT PhD grant.
\end{acknowledgments}


\end{document}